\documentclass[9pt,twocolumn,twoside]{opticajnl}
\journal{opticajournal} 

\setboolean{shortarticle}{false}

\usepackage{bm}

\usepackage{lineno}

\title{Achieving 100$\,$MHz Instantaneous Bandwidth in a Broadband Rydberg Microwave Sensor}

\author[1,3,$\dagger$]{Yuhan Yan}
\author[1,2,$\dagger$]{Jinyin Wan}
\author[1,3]{Xuejie Li}
\author[1,2]{Xing Xia}
\author[1,2]{Haojie Zhao}
\author[1,2]{Binghong Yu}
\author[1,2,5]{Jianliao Deng}
\author[3,4,6]{L. Q. Chen}
\author[1,2,7]{Huadong Cheng}

\affil[1]{Wangzhijiang Innovation Center for Laser, Aerospace Laser Technology and System Department, Shanghai Institute of Optics and Fine Mechanics, Chinese Academy of Sciences, Shanghai 201800, China}
\affil[2]{Center of Materials Science and Optoelectronics Engineering,
	University of Chinese Academy of Sciences, Beijing 100049, China}
\affil[3]{State Key Laboratory of Precision Spectroscopy, Institute of Quantum Science and Precision Measurement, School of Physics, East China Normal University, Shanghai, 200062, China}
\affil[4]{Hefei National Laboratory, Hefei 230088, China}

\affil[$\dagger$]{These authors contributed equally to this work.}
\affil[5]{jldeng@siom.ac.cn}
\affil[6]{lqchen@phy.ecnu.edu.cn}
\affil[7]{chenghd@siom.ac.cn}

\begin{abstract}
Rydberg atoms have attracted considerable attention in recent years as a novel platform for microwave sensing, owing to their unique physical merits: large transition dipole moments between Rydberg levels and broad frequency coverage. As a critical figure of merit for Rydberg microwave sensors, instantaneous bandwidth serves as a key benchmark for evaluating their viability in practical applications. Previous studies on instantaneous bandwidth remain limited to single-frequency operation, with typical demonstrated values of only tens of megahertz, a constraint that hampers the real-world deployment of this sensing technology.
Here, we experimentally achieve an instantaneous bandwidth of over 100$\,$MHz across a broad frequency range of 2.7–20$\,$GHz and realize a sensitivity in the hundreds of nV$\,$cm$^{-1}\,$Hz$^{-1/2}$ range. The physical mechanism lies in the dressed-state coherence and the interference effect between different transition channels. Our work substantially broadens the instantaneous bandwidth of Rydberg microwave sensors and paves the way for their practical deployment in fields such as radar and wireless communications.
\end{abstract}

\setboolean{displaycopyright}{false} 

\begin{document}

\maketitle

\section{Introduction}
Rydberg atoms, owing to their exceptionally large electric dipole moments and broad frequency coverage, have attracted considerable attention in recent years as a novel platform for microwave (MW) sensing \cite{saffman2010quantum,holloway2014broadband,holloway2017atom,holloway2017electric,anderson2020rydberg,meyer2020assessment,meyer2021waveguide, sedlacek2012microwave,sedlacek2013atom,anderson2014two,miller2016radio,anderson2016optical,jing2020atomic,prajapati2021enhancement,liu2022continuous,zhang2022rydberg,liu2022deep,ding2022enhanced,tu2022high,ouyang2023continuous,berweger2023closed,borowka2024continuous,tu2024approaching}. The advent of the superheterodyne measurement paradigm has revolutionized the sensitivity performance of Rydberg MW sensors, delivering orders-of-magnitude enhancements over conventional Autler-Townes splitting detection approaches \cite{sedlacek2012microwave, jing2020atomic}. However, for Rydberg microwave sensors to achieve widespread practical deployment, the fundamental challenge of their narrow instantaneous bandwidth (IB) must be overcome. Consequently, since the inception of the superheterodyne approach, researchers have devoted significant efforts to enhancing the IB, employing techniques including two-beam excitation \cite{hu2023improvement}, probe light frequency combs \cite{dixon2023rydberg, artusio2024increased}, and external magnetic fields \cite{11004473}. In 2024, our group demonstrated that the superheterodyne signal originates from the beat note between the probe field and the sidebands generated by the two six-wave mixing processes \cite{yang2024highly}. By increasing the Rabi frequency of the coupling laser, we achieved an IB of $\pm$10.2$\,$MHz. Subsequently, in 2025, we revealed that the IB of the superheterodyne signal depends on the coupling between dressed states, and realized an IB exceeding 50$\,$MHz via multi-dressed-state engineering \cite{yan2026multidressed}. In 2026, a novel scheme for enhancing IB by introducing an auxiliary MW field is proposed and an IB of $\pm$22.3$\,$MHz is achieved when the coupling light at resonance \cite{yan2026broadbandrydbergatomicmicrowave}. As a result, the IB has been improved from the initial 150$\,$kHz to tens of megahertz. However, the IB reported in these prior studies does not exceed 100$\,$MHz, with performance evaluations limited to a single, narrow frequency band around a chosen transition.

In this work, we experimentally demonstrate a Rydberg MW sensor with operational coverage over the 2.7–20$\,$GHz frequency band. By strategically tuning the coupling laser detuning, the dressed-state coherence and the interference between different transition channels are altered, therefore we simultaneously achieve an IB exceeding 100$\,$MHz and hundreds of nV$\,$cm$^{-1}\,$Hz$^{-1/2}$ sensitivity.
Our work extends the IB of Rydberg MW sensors beyond 100$\,$MHz across a wide frequency range for the first time, demonstrating the substantial application potential of Rydberg MW sensing and advancing its progress toward practical deployment.

\section{Theoretical analysis}
\subsection{The superheterodyne response}
\begin{figure}[t!]
	\centering
	\includegraphics[width=0.65\linewidth]{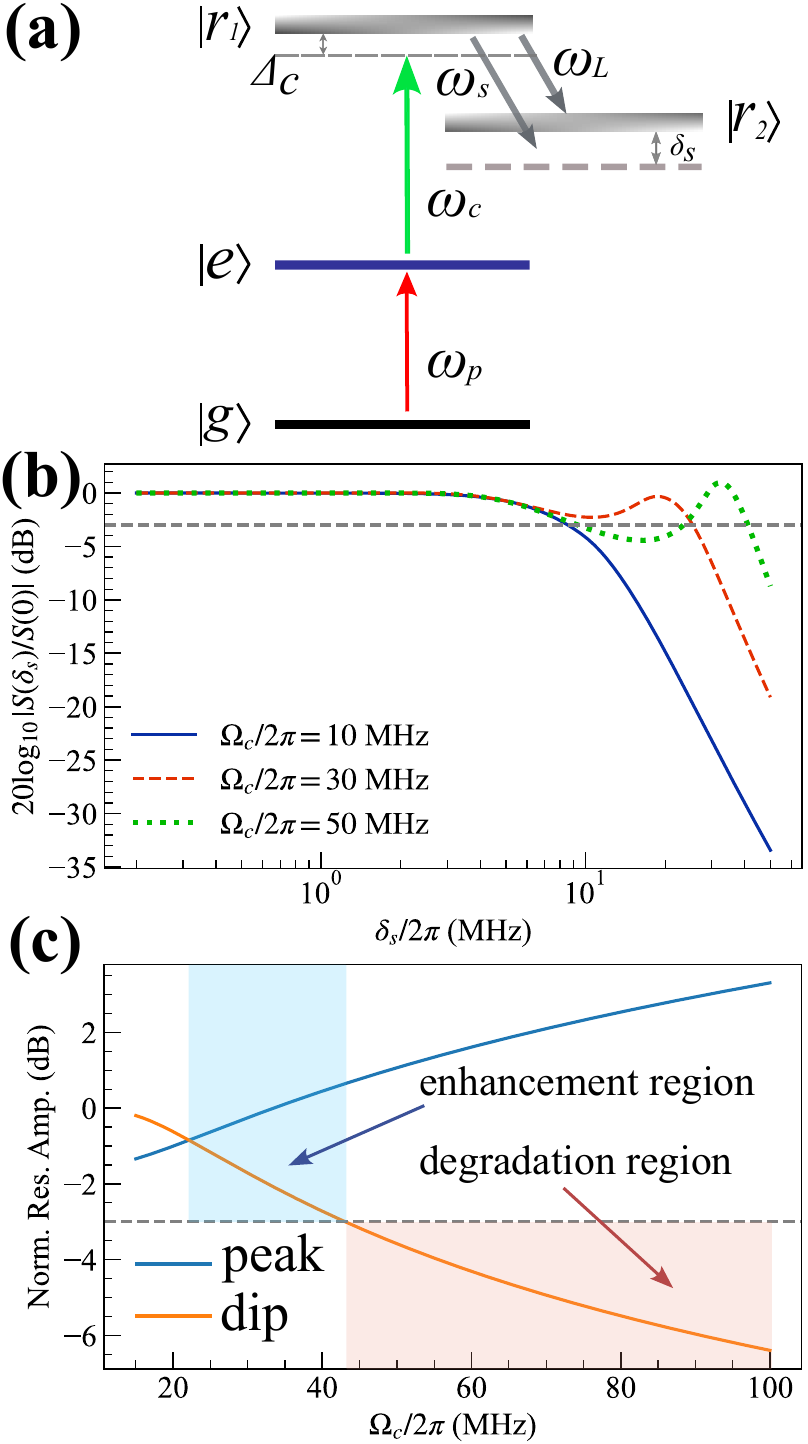}
	\caption{(a) Energy level scheme. $\omega_{i}$: $i=p, c, L, s$ represent the angular frequency of the probe, coupling, local MW, and signal MW fields, respectively.  $|g\rangle$: $|6S_{1/2}, \ F=4\rangle$, $|e\rangle$: $|6P_{3/2}, \ F'=5\rangle$, $|r_1\rangle$: $|nD_{5/2}\rangle$, $|r_2\rangle$: $|(n+1)P_{3/2}\rangle$. (b) The normalized response amplitude curves with different $\Omega_c$. The gray dashed line represents the 3$\,$dB attenuation of the response amplitude. The data is normalized to the response amplitude of $|\delta_s|/2\pi=200\,$kHz. (c) The normalized response amplitude of the peak and dip as a function of $\Omega_c$. 
	}
	\label{fig1}
\end{figure} 
The energy level scheme for a Rydberg superheterodyne MW sensor is shown in Fig. \ref{fig1}. The atoms at the ground state $|g\rangle $ are excited to the Rydberg state $|r_{1}\rangle $ by the probe field (with a frequency $\omega_{p}$ and a Rabi frequency $\Omega_p$) and coupling field (with a frequency $\omega_{c}$ and a Rabi frequency $\Omega_c$). The probe field is resonant on the transition frequency of $\left\vert
g\right\rangle \rightarrow \left\vert e\right\rangle $.  The coupling field continuously drives the atomic transition from $\left\vert
e\right\rangle$ to $\left\vert r_1\right\rangle$.
The local MW field (with a frequency $\omega_{L}$ and a Rabi frequency $\Omega_{L}$) is coupling the Rydberg states $|r_{1}\rangle $ and $|r_{2}\rangle$ and the signal MW field as a perturbation drives the system with a Rabi frequency $\Omega_s$ ($\Omega_s \ll \Omega_L$) and a frequency detuning of $\delta_s=\omega_s-\omega_L$. For such a four-level light-atom interaction system, the Hamiltonian under the rotating wave approximation (RWA) is 
\begin{equation}
	\begin{aligned}
		\hat{H}_R&=-\left[\Delta_p \hat{\sigma}_{ee}+\left(\Delta_p+\Delta_c\right)\hat{\sigma}_{r_{1}r_{1}}+\left(\Delta_p+\Delta_c-\Delta_L\right)\hat{\sigma}_{r_{2}r_{2}}\right] \\
		&+\frac{1}{2}\left[\Omega_p \hat{\sigma}_{ge}+\Omega_c \hat{\sigma}_{er_{1}}+\left(\Omega_L+\Omega_s e^{i\delta_s t}\right) \hat{\sigma}_{r_{1}r_{2}}+H.c.\right]
	\end{aligned}
	\label{eqhalr}
\end{equation}
where the light detuning $\Delta_x\,(x\in\{p,c,L\})$ are defined as
\begin{equation}
	\begin{aligned}
		\Delta_p &=\omega_p-\omega_{ge}\\
		\Delta_c &=\omega_c-\omega_{er_{1}}\\
		\Delta_L &=\omega_L-\omega_{r_{1}r_{2}}
	\end{aligned}
	\label{deltapcl}
\end{equation}
and $\hat{\sigma}_{m n}=|m \rangle \langle n|$ ($m,n\in \{g,e,r_1,r_2\}$) is the atomic transition operator which represents the transition from the atomic state $|n \rangle$ to $|m \rangle$, $\omega_{mn}$ is the resonant frequency between atomic states $|m\rangle$ and $|n\rangle$. In the dissipation system, the dynamics of the atoms interacting with light fields is described by the Lindblad master equation
\begin{equation}
	\partial_t \rho =-i\left[\hat{H}_R,\rho\right]+\hat{L}\left[\rho\right]
	\label{eqlind}
\end{equation}
where $\rho$ is the density matrix of the system and $\hat{L}[\rho]$ is the Lindblad superoperator describing the dissipation. According to the Floquet theory \cite{Eckardt_2015,Rodriguez-Vega_2018,rudner2020floquetengineershandbook}, under the driving of a periodic signal MW field $\Omega_se^{i\delta_st}+H. c.$, the atomic coherence $\rho_{ge}$ can be expressed as $\rho_{ge}=\rho_{ge}^{0}+\rho_{ge}^{+1}e^{i\delta_s t}+\rho_{ge}^{-1}e^{-i\delta_s t}$. Thus, the response amplitude of the superheterodyne, which is the absorption of the probe field influenced by the signal MW field, can be expressed as $|S(\delta_s)|=\left\vert \rho_{ge}^{+1}-\rho_{ge}^{-1*}\right\vert$. The IB is determined from the normalized frequency response curve of $S_N(\delta_s)=20\log_{10}|S(\delta_s)/S(0)|$.  It corresponds to the total frequency span where $S_N(\delta_s)$ drops to --3$\,$dB. By solving Eq. (\ref{eqlind}), we present three superheterodyne frequency response curves in Fig. \ref{fig1}(b), corresponding to different $\Omega_c$ when $\Delta_c=0$. As $\Omega_c$ increases, a dip initially emerges. With a further increase in $\Omega_c$, a gain peak appears in the response curve. When the depth of this dip exceeds 3$\,$dB, the IB is severely degraded. Figure \ref{fig1}(c) illustrates the regimes of IB enhancement and degradation. Beyond a critical threshold of $\Omega_c$, further IB enhancement is inhibited despite the presence of the gain peak, as the dip depth surpasses 3$\,$dB. Therefore, to achieve substantial IB enhancement, the critical first step is to suppress the response dip. To this end, we first analyze the underlying physical origins of the dip and peak features.
\subsection{The dressed-state picture}
The Hamiltonian \(\hat{H}_R\) can be decomposed into two terms: one describes the strong-field interactions involving the probe ($\omega_p$), coupling ($\omega_c$), and local MW ($\omega_L$) field, while the other contains solely the weak signal MW field ($\omega_s$). Driven by the three strong driving fields, the system supports four dressed states, labeled \(|-\rangle\), \(|d\rangle\), \(|u\rangle\), and \(|+\rangle\) in ascending order of eigenenergy. Since the strong driving fields participate in  the formation of the dressed states, tuning the strong field parameters ($\Omega_{p,c,L}$ and $\Delta_{p,c,L}$) modifies the dressed-state distribution--that is, it alters their eigenenergies. Specifically, tuning the Rabi frequency of the strong driving fields ($\Omega_{p,c,L}$) modifies the eigenenergies of the dressed states while preserving their symmetric energy configuration. In contrast, adjusting the detuning of the strong fields ($\Delta_{p,c,L}$) not only alters the dressed-state eigenenergies but also induces a global shift of all energy levels. The direction of this shift depends on whether the detuning is red or blue, which breaks the symmetric distribution of the dressed-state manifold. We now turn to the effect of the weak signal MW field. By separating its contribution to the Hamiltonian $\hat{H}_{R}$ from Eq. (\ref{eqhalr}) and transforming it into the dressed-state representation, we obtain
\begin{equation}
	\hat{H}_{s}=\frac{\Omega_s}{2} e^{i\delta_s t}\sum_{ij} \bm{C}_{ij}\hat{\sigma}_{ij}+ H.c.
	\label{eq1}
\end{equation}
$\hat{\sigma}_{ij}$ is the atomic transition operator between dressed states $|i\rangle$ and $|j\rangle$ ($i, j \in \{-,d,u,+\}$), and  $\bm{C}_{ij}$ (determined by $\Omega_{p,c,L}$ and $\Delta_{p,c,L}$) is the projection coefficient \cite{bracht2023dressed,li2024resonant, liu2006atom}.
Therefore, the signal field is driving transitions between dressed states with Rabi frequency $\Omega_s$ and a frequency detuning $\delta_s$.
According to Ref. \cite{yan2026multidressed}, the response amplitude is governed by the six dressed-state transition channels ($|- \rangle \leftrightarrow |d \rangle$, $|- \rangle \leftrightarrow |u \rangle$, $|- \rangle \leftrightarrow |+ \rangle$, $|d \rangle \leftrightarrow |u \rangle$, $|d \rangle \leftrightarrow |+ \rangle$, $|u \rangle \leftrightarrow |+ \rangle$), which is given by 
\begin{equation}
	\left\vert\tilde{S}(\delta_s)\right\vert=\sqrt{\sum_{k}|\tilde{S}_k(\delta_s)|^2+\sum_{\alpha \beta}|\tilde{S}_\alpha(\delta_s)||\tilde{S}_\beta(\delta_s)|\cos\Delta\phi_{\alpha \beta}}
	\label{eq2}
\end{equation}
\begin{figure}[t!]
	\centering
	\includegraphics[width=1\linewidth]{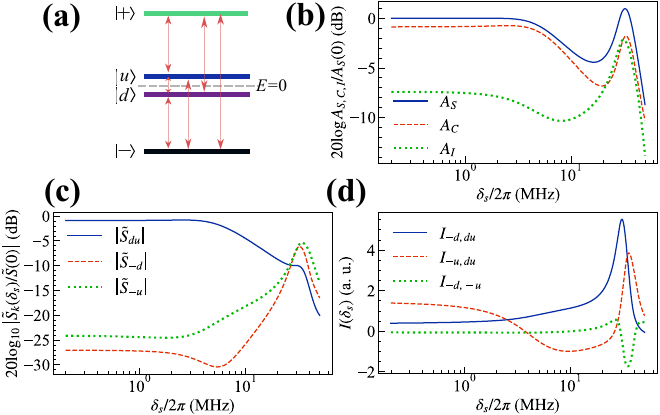}
	\caption{(a) The dressed-state distribution when $\Delta_c=0$. (b) The theoretical calculation of the overall response ($A_S=|\tilde{S}(\delta_s)|$), the contribution of the coherence terms ($A_C=\sum|\tilde{S}_k(\delta_s)|$), and the contribution of the interference terms ($A_I=\sqrt{\sum I_{\alpha \beta}}$). All the theoretical results are normalized to the overall response amplitude $A_S(0)$. (c) The three dominant coherence terms. (d) The three dominant interference terms. All the theoretical results are calculated with $\Omega_p/2\pi=5\,$MHz and $\Omega_c/2\pi=50\,$MHz.
	}
	\label{fig2}
\end{figure} 
where $k, \alpha, \beta \in \left\{| i \rangle \leftrightarrow | j \rangle\right\}$ represent the different dressed-state transition channels, $|\tilde{S}_{k, \alpha, \beta}(\delta_s)|$ is the oscillation amplitude in the imaginary part of the coherence  between dressed states $ | i \rangle $ and $| j \rangle$, $\Delta \phi_{\alpha \beta}$ is the phase difference between $\tilde{S}_{\alpha}(\delta_s)$ and $\tilde{S}_{\beta}(\delta_s)$. Accordingly, the response of the superheterodyne signal is governed by the dressed-state coherence $|\tilde{S}_{k}|$ and the interference between dressed-state transition channels $I_{\alpha\beta}(\delta_s)=|\tilde{S}_{\alpha}(\delta_s)| |\tilde{S}_{\beta}(\delta_s)|\cos\Delta\phi_{\alpha \beta}$.
\begin{figure}[t!]
	\centering
	\includegraphics[width=0.8\linewidth]{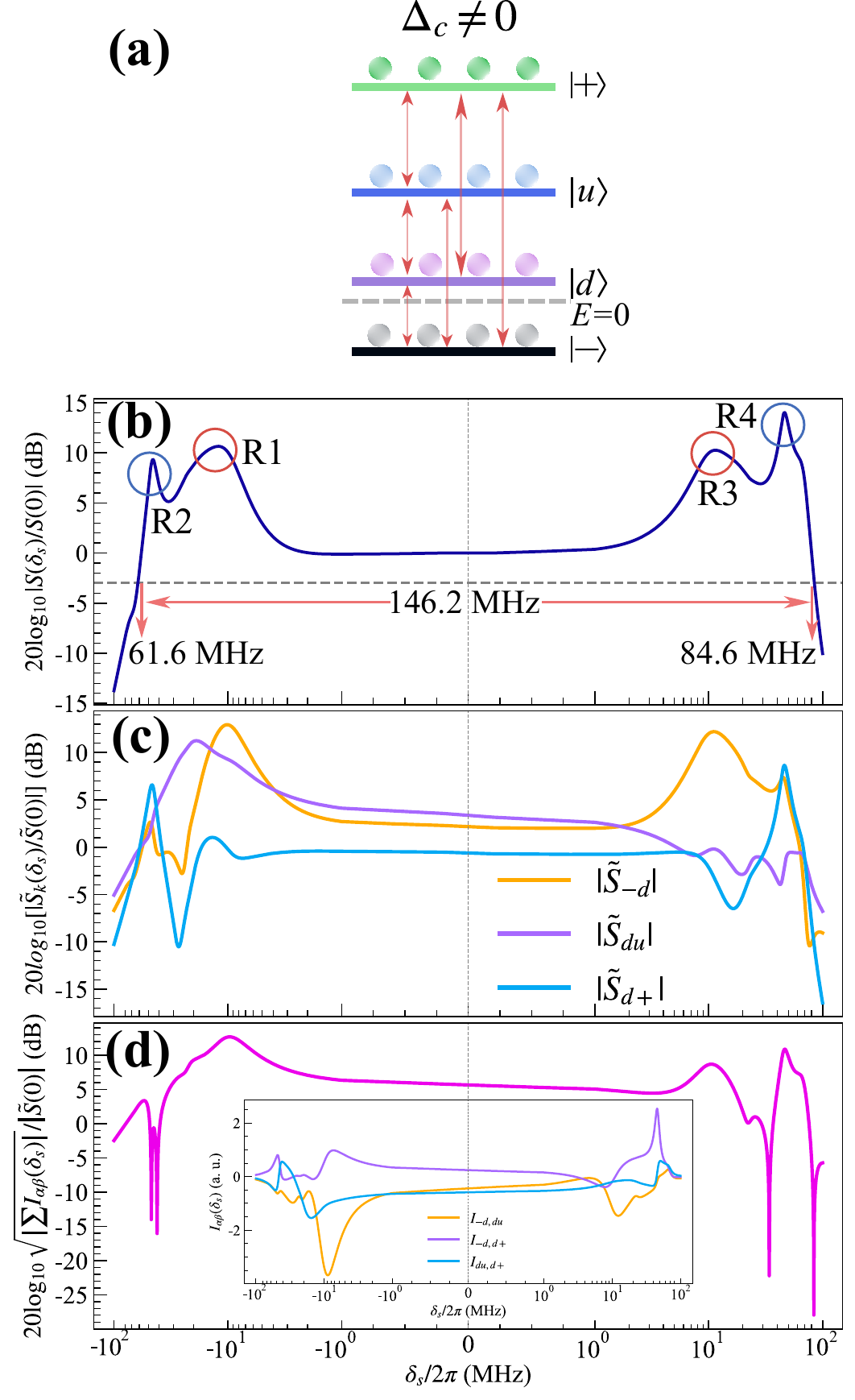}
	\caption{(a) The dressed-state distribution when $\Delta_c \ne 0$. (b) The theoretical normalized response amplitude curve of the overall response. (c) The three dominant coherence terms. (d) The contribution of the interference effect to the overall response. The inset represents the three dominant interference terms. All the theoretical results are calculated with $\Omega_p/2\pi=5\,$MHz, $\Omega_c/2\pi=50\,$MHz, and $\Delta_c/2\pi=-30\,$MHz.
	}
	\label{fig3}
\end{figure} 

For \(\Delta_c = 0\), the corresponding physical picture of the dressed states is illustrated in Fig. \ref{fig2}(a), where the dressed-state energy levels take on a symmetric configuration. We calculate the contributions of the coherence terms and interference terms to the superheterodyne signal by Eq. (\ref{eq2}), with the results presented in Fig. \ref{fig2}(b-d). When $|\delta_s|$ is tuned from zero, the signal MW field first couples the dressed states \(|d\rangle\) and \(|u\rangle\). The coherence between \(|d\rangle\) and \(|u\rangle\) is established first, such that \(|\tilde{S}_{du}|\) dominates the superheterodyne signal in the low-$|\delta_s|$ regime.
However, under the strong coupling condition, \(\omega_{-d} \gg \omega_{du}\) ($\omega_{ij}$ is the resonant frequency of the dressed states $|i\rangle$ and $|j\rangle$), as $|\delta_s|$ is tuned away from \(\omega_{du}\), the coherence between \(|d\rangle\) and \(|u\rangle\) decays, while coherence between other pairs of dressed states has not yet been established, as illustrated in Fig. \ref{fig2}(c), this gives rise to the dip in the superheterodyne signal. When $|\delta_s|$ approaches $\omega_{-d}$, since $\omega_{-d}=\omega_{u+}\approx\omega_{d+}=\omega_{-u}$, the coherence is fully established between all other pairs of dressed states, and our calculation results in Fig. \ref{fig2}(b) show that the overall interference of transitions among the dressed states is constructive. Therefore, the enhanced coherence between dressed states and the constructive interference of transitions collectively give rise to the formation of the peak. Based on the analysis, a method capable of compensating for the coherence degradation between dressed states is required to prevent the response dip from becoming excessively deep. Accordingly, it is essential to introduce coupling field detuning to redistribute the dressed states.
\begin{figure*}[t!]
	\centering
	\includegraphics[width=0.85\linewidth]{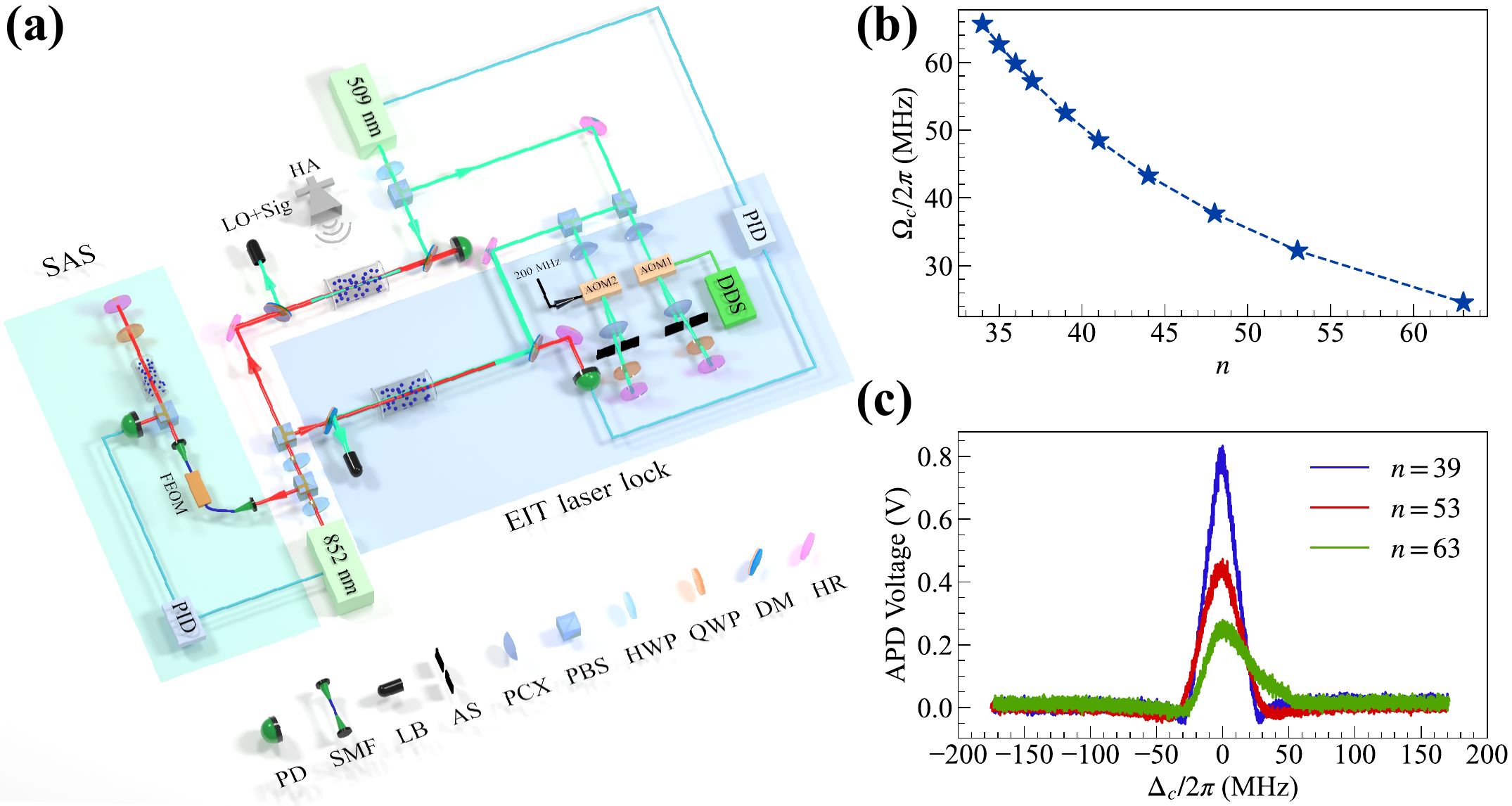}
	\caption{(a) Experimental setup. HR: high-reflection mirror; DM: dichroic mirror; HWP: half wave plate; QWP: quarter wave plate; BS: beam splitter; PBS: polarization beam splitter; PCX: plano-convex lens; AS: aperture stop; LB: laser block; SMF: single-mode fiber; FEOM: fiber electro-optic modulator;AOM: acousto-optic modulator; PD: photodetector; PID: proportional-integral-derivative feedback; DDS: direct digital synthesizer; SAS: saturated absorption spectroscopy; EIT: electromagnetically induced transparency; LO: local MW field; Sig: signal MW field; HA: horn antenna. (b) The peak Rabi frequency of the coupling field with different principle quantum number $n$. The power of the coupling beam is maintained at 400$\,$mW.
		(c) The EIT signals with different principle quantum numbers.
	}
	\label{fig4}
\end{figure*}

The dressed-state distribution when $\Delta_c \ne 0$ is shown in Fig. \ref{fig3}(a). The introduction of $\Delta_c$ disrupts the symmetric configuration of the dressed states, enabling controllable tuning of both the eigenenergies and relative spacings of the dressed-state manifold.
We calculated the superheterodyne response curve  with $\Omega_p/2\pi=5\,$MHz and $\Omega_c/2\pi=30\,$MHz, which is presented in Fig. \ref{fig3}(b). The dip that degrades the IB is eliminated, and multiple gain peaks emerge. This yields a substantial enhancement of the IB to over 100$\,$MHz. The IB enhancement arises fundamentally from the modification of dressed-state coherence and the interference effect when $\Delta_c \ne0$. The three dominant coherence terms are shown in Fig. \ref{fig3}(c), and the contribution of the interference effect to the superheterodyne signal is also shown.
When \(\delta_s < 0\), both $|\tilde{S}_{du}|$ and $|\tilde{S}_{-d}|$ exhibit an upward trend, and when \(\delta_s > 0\), the increasing trend of $|\tilde{S}_{-d}|$ compensates for the decrease in $|\tilde{S}_{du}|$. Accordingly, peaks emerge in the \(R_1\) and \(R_3\) regimes respectively as $|\tilde{S}_{-d}|$ reaches its peak position although there is a destructive interference effect near this $|\delta_s|$ position. As $|\delta_s|$ increases further, $|\tilde{S}_{d+}|$ begins to rise and compensates for the attenuation of $|\tilde{S}_{-d}|$ and $|\tilde{S}_{du}|$, combining the interference effect shown in the inset of Fig. \ref{fig3}(d), resulting in the emergence of peaks in the \(R_2\) and \(R_4\) regimes when $|\tilde{S}_{d+}|$ is reaching its peak. 
Therefore, by tuning the coupling detuning $\Delta_c$, we modulate the dressed-state coherence and the interference between different transition channels, giving rise to resonance peaks between dressed-state pairs and thereby significantly extending the IB. This physical picture accounts for our experimental demonstration of over 100$\,$MHz IB at multiple frequency points across a wide frequency range of 2.7-20$\,$GHz. 

\begin{figure*}[t!]
	\centering
	\includegraphics[width=1\linewidth]{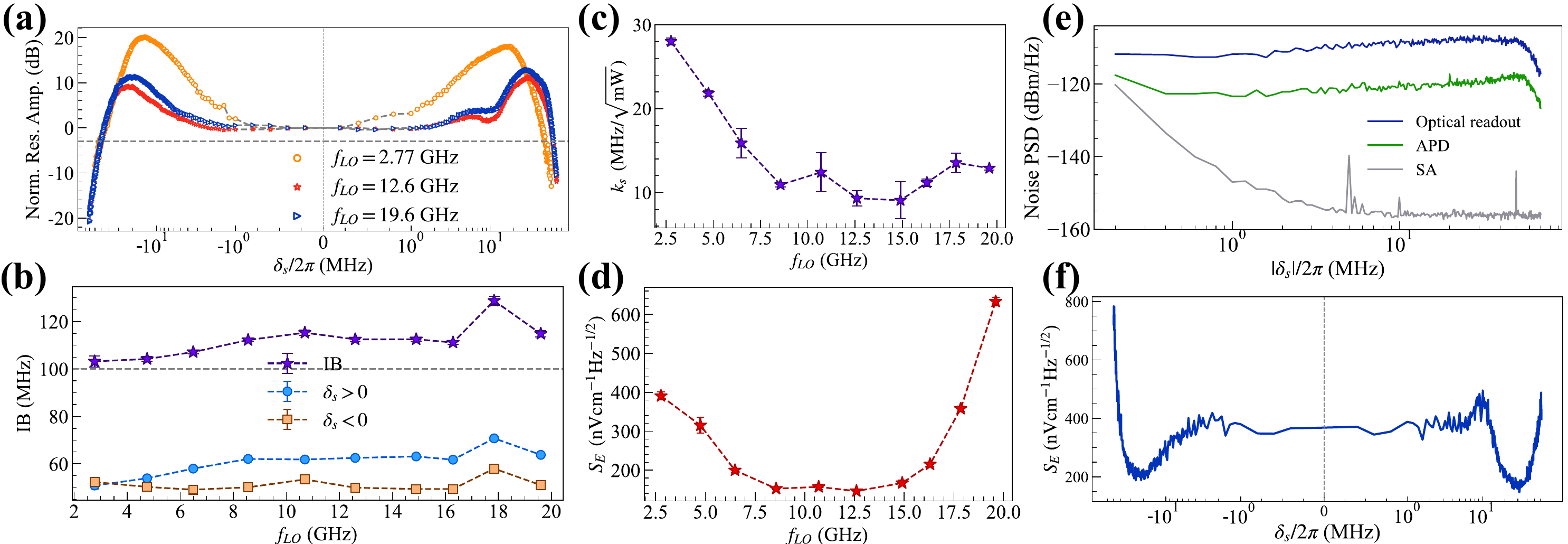}
	\caption{(a) The experimental normalized response amplitude when $f_{LO}=2.77\,$GHz,  $f_{LO}=12.6\,$GHz and $f_{LO}=19.6\,$GHz. The gray dashed line represents the 3$\,$dB attenuation of the response amplitude at $|\delta_s|/2\pi=200\,$kHz. (b) The IB performance at multiple frequency point.
		During the IB measured process, the signal power was kept at --45$\,$dBm. (c) The slope of the linear fitting of the AT splitting and the signal MW power. (d) The sensitivity of the Rydberg MW sensor at multiple frequency point. The peak Rabi frequency of the probe field $\Omega_p/2\pi=15.69\,$MHz was maintained with all the measurements. The error bars shown in the figures are obtained from five independent measurements. (e) The measured noise power spectral density spectrum (PSD) of the optical readout, APD, and SA. (f) The noise limited sensitivity at $f_{LO}=12.6\,$GHz.
	}
	\label{fig5}
\end{figure*}
\section{Experimental setup}
The experimental setup is shown in Fig. \ref{fig4}(a). The core is a Cs vapor cell in the room temperature of 21$\,$℃. Our experimental setup employs a probe beam generated by an 852$\,$nm laser and a coupling beam produced by a frequency-tunable
509$\,$nm laser. This 509$\,$nm laser provides a sufficiently broad tuning range to excite Cs atoms to Rydberg states with principal quantum numbers ranging from $n=34$ to $n=63$,  which covers the 2.7–20$\,$GHz transition frequency band between the $nD_{5/2}$ and $(n+1)P_{3/2}$ Rydberg states. The 852$\,$nm laser is frequency locked to the $|6S_{1/2},F=4\rangle \rightarrow |6P_{3/2},F'=5\rangle $ transition by the saturated absorption spectroscopy (SAS). The 509$\,$nm laser is frequency locked to the transmission peak of the electromagnetically induced transparency (EIT) signal, corresponding to the resonant condition between the excited state $|e\rangle$ and the Rydberg state $|r_1\rangle$.
Before entering the Cs vapor cell used for frequency stabilization, the 509$\,$nm laser is frequency-shifted using two acousto-optic modulators (AOMs) with a center frequency of 200$\,$MHz.  The optical path is specifically designed such that the coupling beam undergoes double-pass transmission through the negative first-order diffraction of AOM1 and the positive first-order diffraction of AOM2. This double-pass AOM arrangement enables continuously tuning of the coupling laser frequency, while simultaneously ensuring that the propagation direction of the output coupling beam. The probe power is 3$\,\mu$W with a $1/e^2$ waist radius of 80$\,\mu$m at the center of vapor cell, yielding a peak Rabi frequency $\Omega_p/2\pi=15.69\,$MHz. The incident power of the coupling beam is maintained at 400$\,$mW with a $1/e^2$ waist radius of 85$\,\mu$m at the center of vapor cell. Since the transition dipole moment between $|e\rangle$ and $r_1\rangle$ varies for different Rydberg states $|r_1\rangle$, the Rabi frequency of the coupling beam differs at different frequency points. The corresponding peak Rabi frequencies at each frequency point are shown in Fig. \ref{fig4}(b).
The probe and coupling beams are maintained in a counter-propagating configuration throughout the experiment to reduce the Doppler effect.
The local MW and the signal MW are combined via a power combiner and then emitted to the 5-cm vapor cell through a horn antenna. The horn antenna is positioned more than 50$\,$cm away from the vapor cell to ensure uniform MW illumination, and the vapor cell is surrounded by microwave anechoic absorbers to minimize the impact of MW reflections from metallic surfaces on the measured signal. All the optical and MW fields are vertically polarized. The superheterodyne signal is detected by an APD (Thorlabs APD130A) with a detection bandwidth of 50$\,$MHz and acquired by a spectrum analyzer (Agilent N9020A). All the signal generators and spectrum analyzer are synchronized to a 10$\,$MHz reference signal source.
\section{Results and discussion}
The electromagnetically induced transparency signals with different principle quantum numbers are shown in Fig. \ref{fig4}(c). As the principal quantum number increases, the Rabi frequency of the coupling $\Omega_c$ decreases progressively. This leads to a reduced atomic population in the Rydberg state, which in turn lowers the signal-to-noise ratio (SNR) of the EIT signal.
We measured the frequency response curves of the superheterodyne signal for the Rydberg states with principal quantum number from $n = 34$ to $n = 63$. Building on the preceding physical analysis of IB tuning, two prerequisites must be satisfied to achieve a high IB exceeding 100$\,$MHz: suppressing the response dip that degrades IB performance, and generating gain peaks arising from dressed-state resonances to extend the IB. Accordingly, the coupling detuning $\Delta_c$ must be carefully adjusted and optimize $\Omega_L$ at each operating frequency point to achieve the maximum attainable IB.
Figure \ref{fig4}(a) presents the frequency response curves measured at two frequency points of 2.77$\,$GHz and 19.6$\,$GHz, where the IB clearly exceeds 100$\,$MHz in both cases. Figure \ref{fig4}(b) shows the measured IB at various local MW frequencies. It can be observed that the IB exceeds 100$\,$MHz across all measured frequency points, and surpasses 50$\,$MHz on both the \(\delta_s<0\) and \(\delta_s>0\) sides.

Experimentally, the bandwidth-normalized sensitivity is calculated from the minimum detectable power ($P_{min}$) corresponding to a signal-to-noise ratio (SNR) of unity, as given by $S_E=\eta_s \sqrt{10^{P_{min}/10}/f_{RBW}}$,
where $\eta_s=k_sh/\mu_{r_1 r_2}$, $h$ is the Planck constant, $\mu_{r_1  r_2}$ is the transition dipole moment of the Rydberg states $|r_1\rangle$ and $|r_2\rangle$, $f_{RBW}$ is the resolution bandwidth of the spectrum analyzer, and $k_s$ is the slope of the linear fitting of the Autler-Townes (AT) splitting and the signal MW power.
The measured $k_s$ at different local MW frequency points are shown in Fig. \ref{fig5}(a). The results of achievable sensitivity at each frequency point are shown in Fig. \ref{fig5}(b). 
It is observed that the sensitivity reaches 147$\,$nV$\,$cm$^{-1}\,$Hz$^{-1/2}$ when $f_{LO}=12.6\,$GHz. For $f_{LO}<12.6\,$GHz, the transition dipole moment for the \(|e\rangle\to|r_2\rangle\) transition is smaller, which reduces the Rabi frequency of the coupling field and lowers the atomic population in Rydberg state \(|r_1\rangle\). This diminished Rydberg population in turn degrades low-frequency sensitivity. For $f_{LO}>12.6\,$GHz, the transition dipole moment of the \(|r_1\rangle \to |r_2\rangle\) transition decreases progressively, which weakens the coupling between the microwave fields and the Rydberg states, thereby degrades the sensitivity. Figure \ref{fig5}(e) presents the noise power spectral density (PSD) of our system. The sensitivity at $f_{LO}=12.6\,$GHz in the range of 3$\,$dB IB is given in Fig. \ref{fig5}(f). Across the full 3-dB IB range, the sensitivity consistently remains better than 800$\,$nV$\,$cm$^{-1}\,$Hz$^{-1/2}$.

\section{Conclusion}
In conclusion, we theoretically elucidate the physical mechanism by which introducing coupling  detuning enhances the instantaneous bandwidth and demonstrate a multi-frequency Rydberg microwave sensor, with the bandwidth-improvement scheme both theoretically analyzed and experimentally implemented. By tuning the coupling laser detuning, we demonstrate a Rydberg microwave sensor operating over the wide frequency band of 2.7–20$\,$GHz, and an instantaneous bandwidth exceeding 100$\,$MHz and hundreds of nV$\,$cm$^{-1}\,$Hz$^{-1/2}$ sensitivity are attained at all operating frequency points which can be improved further by increasing the power of coupling laser. Our work not only verifies the underlying physical mechanism for the tunability of instantaneous bandwidth--governed by the dressed-state coherence and interference between different transition channels--but also substantially broadens the achievable instantaneous bandwidth of Rydberg microwave sensors, greatly advancing the practical applications of Rydberg atoms in microwave metrology.








\begin{backmatter}
\bmsection{Funding}
National Key Research and Development Program of China (2024YFA1409404), the National Natural Science Foundation of China (Grants No. 12174409, No. 12304294, No. U23A2075, and No. 12274132), the Fundamental Research Funds for the Central Universities and Shanghai Municipal Education Commission (202101070008E00099).

\bmsection{Acknowledgment}
Y. Y and J. W demonstrated the experimental setup, performed the measurements and analyzed data. The theoretical analysis and numerical simulations are conducted by Y. Y. All other authors discussed the experimental results. X. L., X. X., H. Z., and B. Y. gave assistance for collecting data. The manuscript was written by Y. Y. J. D. ,  L. C.,  and H. C.  supervised the study and suggested the writing of the manuscript. 

\bmsection{Disclosures}
The authors declare no conflicts of interest.

\bmsection{Data availability} Data underlying the results presented in this paper are not publicly available at this time but may be obtained from the authors upon reasonable request.
\end{backmatter}

\bibliography{100MHz}




\end{document}